\documentclass[runningheads,a4paper]{llncs}
\usepackage{etex}
\pdfoutput=1

\usepackage[linesnumbered,lined,boxed,commentsnumbered,ruled]{algorithm2e}

\SetAlFnt{\small}
\SetAlCapFnt{\small}
\SetAlCapNameFnt{\small}
\SetAlCapHSkip{0pt}
\IncMargin{-\parindent}
\usepackage{ textcomp }
\usepackage{ comment }
\usepackage[color=yellow!60,textsize=footnotesize,obeyDraft,draft]{todonotes}
\usepackage{colortbl}
\usepackage{booktabs}
\usepackage{capt-of}
\usepackage{makecell}
\usepackage{textcomp }
\usepackage{multirow}
\usepackage{listings}
\usepackage{graphicx}
\usepackage{colortbl}
\usepackage{booktabs}
\usepackage{amsmath}
\usepackage[font=footnotesize,labelfont=bf]{subcaption}
\usepackage[hidelinks]{hyperref}
\usepackage[capitalize]{cleveref}
\usepackage{tikz}
\usepackage{pgfplots}
\usepackage[para,online,flushleft]{threeparttable}
\usepackage{multirow}
\usepackage{wasysym}
\usepackage{amssymb}
\usepackage{enumitem}
\usetikzlibrary{arrows}
\usepackage{subcaption}
\usepackage{breakcites}
\usepackage{breakurl}
\usepackage[hidelinks]{hyperref}
\hypersetup{breaklinks=true}
\newcounter{mnotei}
\newcolumntype{L}[1]{>{\raggedright\let\newline\\\arraybackslash\hspace{0pt}}m{#1}}
\newcolumntype{C}[1]{>{\centering\let\newline\\\arraybackslash\hspace{0pt}}m{#1}}
\newcolumntype{R}[1]{>{\raggedleft\let\newline\\\arraybackslash\hspace{0pt}}m{#1}}

\newcommand{\includegraphicsmaybe}[2]{
    \IfFileExists{#2}{\includegraphics[#1]{#2}}{
    \detokenize{File #2 is missing, maybe you need to run plots.py?}
}}

\begin{document}
\mainmatter

\title{A Comprehensive and Accurate Energy Model for Arm's Cortex-M0 Processor}

\titlerunning{An Accurate Energy Model for Arm's Cortex-M0 Processor}

\author{Kyriakos Georgiou, Zbigniew Chamski, Kris Nikov, Kerstin Eder}

\authorrunning{K. Georgiou et al.}
\institute{University of Bristol, UK}
\tocauthor{Authors' Instructions}
\maketitle

\makeatletter
\renewcommand\subsubsection{\@startsection{subsubsection}{3}{\z@}%
                       {-18\p@ \@plus -4\p@ \@minus -4\p@}%
                       {4\p@ \@plus 2\p@ \@minus 2\p@}%
                       {\normalfont\normalsize\bfseries\boldmath
                        \rightskip=\z@ \@plus 8em\pretolerance=10000 }}
\makeatother

\begin{abstract}
Energy modeling can enable energy-aware software development and assist the developer in meeting an application's energy budget. Although many energy models for embedded processors exist, most do not account for processor-specific configurations, neither are they suitable for static energy consumption estimation. This paper introduces a comprehensive energy model for Arm's Cortex-M0 processor, ready to support energy-aware development of edge computing applications using either profiling- or static-analysis-based energy consumption estimation. The model accounts for the Frequency, PreFetch, and WaitState processor configurations which all have a significant impact on the execution time and energy consumption of edge computing applications. All models have a prediction error of less than 5\%.
\end{abstract}

\section{Introduction}
\label{sec:intro}

One trillion new Internet of Things (IoT) devices is predicted to reach the market by 2035~\cite{ARM_trillion_iot_devices}. These devices would be generating an unprecedented amount of data that would need to be pushed to the cloud for storing and processing~\cite{IDC_iot_data_2025}. Edge computing, however, has enabled a degree of processing at the data-source that avoids the need for transmitting all collected data to the cloud. Although this can significantly reduce response time and bandwidth requirements, it results in increased resource requirements from edge devices, such as processing power and energy. 

Typically, IoT devices are not part of a power grid but rather are scattered in the environment and powered by limited energy sources, such as batteries or energy harvesting. Thus, IoT devices are mostly based on small embedded processors with a tiny energy footprint, such as the Arm Cortex-M0. This kind of processor is inherently limited in processing power, making edge computing challenging. Developers must apply extreme optimizations to trim down the processing time, memory, and energy consumption of algorithms to enable their execution on small embedded devices. A trending example of such optimization is the streaming down of traditional machine learning algorithms to enable their execution on tiny IoT devices~\cite{ARM_Cortex_M_machine_learning_white_paper}. 

The burden now lies with the software engineers to develop edge computing applications that can fit on the limited memory of the IoT embedded devices, execute within reasonable timeframes, and run within the available energy budget. Balancing the resource usage of an IoT application is a challenging act and typically a manual trial-and-error process that costs significant development time. Thus, practical methodologies that enable resource-aware software development will significantly help to tackle the challenges associated with the development of edge computing applications. 

Execution time and code size are easy to measure and well understood by the typical software developer. On the contrary, energy consumption information is not readily accessible, and something most software developers never had to account for. For edge computing, however, energy consumption feedback during the applications' development cycle is at least equally important as execution time and code size. Even when the energy consumption is directly proportional to time, energy consumption figures are still needed to ensure that the application's available energy budget is met~\cite{Eder:2016:ENTRA,GeorgiouK:2020,Georgiou:2018:SCOPES}. Thus, development tools need to enable such feedback~\cite{theIoTChallenge}.

Hardware measurements are the most accurate way of acquiring a program's energy consumption information, but they are not broadly supported by the hardware vendors and not within the know-how of typical software developers. Energy modelling and the integration of energy models into the development toolchains can solve both of these issues \cite{theIoTChallenge}. Once an accurate energy model has been developed for a particular platform, it can be integrated into a toolchain to allow for energy estimations with each compilation. 

The literature offers a plethora of energy consumption models for embedded processors \cite{Penolazzi:2009,Brandolese2011,Kerrison:2015,Georgiou:2017,Yassin2020}. For an energy model to be useful to the software developer, it must be able to convey energy consumption information at the source-code level. Thus, Instruction-Set-Architecture-based (ISA) energy models \cite{Tiwari1996} became the most popular because modeling at the ISA level allows for attributing energy costs to software components, such as ISA Control Flow Graph (CFG) basic blocks. Therefore, ISA-based energy models can be utilized by compiler autotuning techniques to discover energy targeted compiler optimizations \cite{GeorgiouK:2020,Georgiou:2018:SCOPES}, and by static analysis tools \cite{Georgiou:2017,DBLP:journals/corr/Georgiou15} that estimate the energy consumption of programs. 

Although ISA-based energy modeling approaches have benefits, extracting such models is time-consuming and challenging. It requires devising often complex energy measuring procedures to capture the energy consumption of each instruction in the ISA. Typically, an instruction is executed in a tight loop while measuring the power dissipated together with the execution time. For instructions that can not be measured within such a loop, for example, branch instructions, regression analysis is needed to capture their energy consumption. 

On the other hand, energy modeling using Performance Monitoring Counters (PMCs), also named event counters, is a simpler approach than ISA-level modeling. It requires measuring the energy consumption of representative programs, collecting  execution statistics from PMCs and then deducting energy consumption coefficients for each counter via regression analysis \cite{PMU_estimation:2001}.

This paper demonstrates how to build PMC-based models for multiple embedded processor configurations. The models can be used to attribute energy costs to software components and facilitate both profiling-based and static-analysis-based energy consumption estimation, similar to ISA-based models. Our main contributions are:

\begin{enumerate}
\item Because many deeply embedded processors used for IoT devices, such as the ARM Cortex-M0, do not support hardware PMCs, we customized an open-source Instruction Set Simulator (ISS) of the Arm \texttt{thumb} ISA, namely the \texttt{Thumbulator} \cite{Thumbulator}, to produce accurate execution statistics needed for a PMC-style energy model.
\item We identified a set of PMCs that are both statically predictable at ISA basic block level, and at the same time offer a low energy consumption estimation error (a Mean Absolute Percentage Error (MAPE) of less than 5\%).
\end{enumerate}

Only a few of the existing ISA models for deeply embedded processors are parameterized by hardware configurations, but only cover the processor's frequency and voltage \cite{Yassin2020}. Other processor configurations, such as instruction buffer configurations, are equally important as they have a major impact on both the execution time and the energy consumption of an embedded processor. 

\begin{enumerate}\setcounter{enumi}{2}
\item Thus, we enhanced \texttt{Thumbulator} to include such configurations for the STM-32F0xx family of processors \cite{STMF32F0xx}. We tracked the use of the instruction PreFetch buffer (ON/OFF) which aims at increasing the efficiency of instruction fetching, and the number of WaitStates (0/1) required to correctly perform read operations from Flash memory \cite{STMFO_technical}. Our energy models include all the permitted combinations of the selected configurations for a set of commonly used processor frequencies: 20, 24, and 48 MHz. These energy models allow software developers to select the configuration that will meet their application's energy consumption and execution time goals. They can also potentially be used to assess the application's risk of exposure to side-channel attacks (see \Cref{subsec:security}).
\end{enumerate}

\section{Energy Modeling Methodology}
\label{subs:OptRealApps}

\begin{table}[!htb]
\centering
\footnotesize
\begin{tabular}{|c|l|}
\hline
\multicolumn{1}{|c|}{\textbf{Counter}} & \multicolumn{1}{c|}{\textbf{Description}} \\ \hline
$C_{\text{1}}$ & Executed instructions (no Muls) \\ \hline
$C_{\text{2}}$ & Multiplication instructions - Muls\\ \hline
$C_{\text{3}}$ & Taken branches \\ \hline
$C_{\text{4}}$ & RAM data reads \\ \hline
$C_{\text{5}}$ & RAM writes \\ \hline
$C_{\text{6}}$ & Flash data reads \\ \hline
\end{tabular}
\caption{Statically predictable PMCs for energy-modeling.}
\label{tab:Cortex_M0_counters}
\end{table}

Two sets of benchmarks were used for model characterization and validation. First, the BEEBS benchmark suite~\cite{BEEBS:arXiv}; an open-source embedded-system benchmark suite designed for exploring the performance and energy consumption characteristics of embedded architectures. BEEBS supports the MAGEEC open-source energy measurement framework \cite{MageecMeasuring}. The framework provides triggers to start and end the measurements and a calibration factor for each benchmark. This ensures that the benchmark is repeatedly executed in a loop until an adequate sampling number is achieved, and thus, the measurements can be trusted. 76 out of the 88 BEEBS benchmarks have been used. The remaining twelve do not fit in the available memory of our STM32F051 target chip. The second set of benchmarks is based on an industrial edge computing application~\cite{D4.3_TeamPlay}, developed by Irida Labs~\cite{IRIDA_LABS}. The application uses a Convolutional Neural Network (CNN) and implements a smart car-parking monitoring system that can monitor, in real-time, a car parking lot with multiple parking slots to determine whether a slot is occupied or not. The different layers of the CNN, namely convolutional, MaxPool, and Full-Connected, were isolated and configured with different hyper-parameters and optimizations, resulting in 154 distinct benchmarks~\cite{D4.4_TeamPlay}. The MAGEEC energy measurement framework was adapted to support the CNN benchmarks. Overall, a total of 230 benchmarks were used for the training and validation of our energy model. This number goes significantly beyond the average number of used benchmarks reported for existing energy models of embedded processors, ranging between 10-20 benchmarks (see \cite{ke_2015}, pages 22-23, Table 5.1).

\begin{table*}[!htb]
\centering
\resizebox{\textwidth}{!}{%
\begin{tabular}{|l|l|c|c|}
\hline
\rowcolor[HTML]{C0C0C0} 
\multicolumn{1}{|c|}{\cellcolor[HTML]{C0C0C0}\textbf{Hardware Configuration}} & \multicolumn{1}{c|}{\cellcolor[HTML]{C0C0C0}\textbf{Energy Consumption Model [nJ]}} & \multicolumn{1}{c|}{\cellcolor[HTML]{C0C0C0}\textbf{MAPE [\%]}} & \multicolumn{1}{c|}{\cellcolor[HTML]{C0C0C0}\textbf{RESD [\%]}} \\ \hline
[20, OFF, 0] & \begin{tabular}[c]{@{}l@{}} $\text{E} = 0.964258\times{C_{\text{1}}} + 1.652455\times{C_{\text{2}}} + 2.091986\times{C_{\text{3}}} + 1.109833\times{C_{\text{4}}} + 0.650563\times{C_{\text{5}}} + 0.633621\times{C_{\text{6}}}$ \end{tabular} & 2.80 & 3.60 \\ \hline
[20, OFF, 1] & \begin{tabular}[c]{@{}l@{}} $\text{E} = 1.282474\times{C_{\text{1}}} + 2.110668\times{C_{\text{2}}} + 2.191545\times{C_{\text{3}}} + 1.185609\times{C_{\text{4}}} + 0.416602\times{C_{\text{5}}} + 1.178991\times{C_{\text{6}}}$ \end{tabular} & 2.97 & 3.60 \\ \hline
[20, ON, 0] & \begin{tabular}[c]{@{}l@{}} $\text{E} = 1.003378\times{C_{\text{1}}} + 1.885309\times{C_{\text{2}}} + 1.802974\times{C_{\text{3}}} + 1.122833\times{C_{\text{4}}} + 0.849223\times{C_{\text{5}}} + 0.475831\times{C_{\text{6}}}$ \end{tabular} & 2.86 & 3.53 \\ \hline
[20, ON, 1] & \begin{tabular}[c]{@{}l@{}} $\text{E} = 0.895879\times{C_{\text{1}}} + 2.185851\times{C_{\text{2}}} + 2.001178\times{C_{\text{3}}} + 1.493364\times{C_{\text{4}}} + 1.076354\times{C_{\text{5}}} + 1.573758\times{C_{\text{6}}}$ \end{tabular} & 3.68 & 4.61 \\ \hline
[24, OFF, 0] & \begin{tabular}[c]{@{}l@{}} $\text{E} = 0.959172\times{C_{\text{1}}} + 1.888565\times{C_{\text{2}}} + 1.357556\times{C_{\text{3}}} + 1.089427\times{C_{\text{4}}} + 0.993145\times{C_{\text{5}}} + 0.562952\times{C_{\text{6}}}$ \end{tabular} & 3.22 & 3.63 \\ \hline
[24, OFF, 1] & \begin{tabular}[c]{@{}l@{}} $\text{E} = 1.178558\times{C_{\text{1}}} + 2.540429\times{C_{\text{2}}} + 2.042475\times{C_{\text{3}}} + 1.190892\times{C_{\text{4}}} + 0.979651\times{C_{\text{5}}} + 0.891088\times{C_{\text{6}}}$ \end{tabular} & 3.16 & 3.90 \\ \hline
[24, ON, 0] & \begin{tabular}[c]{@{}l@{}} $\text{E} = 0.985415\times{C_{\text{1}}} + 1.933276\times{C_{\text{2}}} + 1.448160\times{C_{\text{3}}} + 1.075671\times{C_{\text{4}}} + 1.011891\times{C_{\text{5}}} + 0.617510\times{C_{\text{6}}}$ \end{tabular} & 3.36 & 3.88 \\ \hline
[24, ON, 1] & \begin{tabular}[c]{@{}l@{}} $\text{E} = 0.883755\times{C_{\text{1}}} + 2.156046\times{C_{\text{2}}} + 1.633465\times{C_{\text{3}}} + 1.436556\times{C_{\text{4}}} + 1.152560\times{C_{\text{5}}} + 1.455166\times{C_{\text{6}}}$ \end{tabular} & 4.15 & 5.02 \\ \hline
[48, OFF, 1] & \begin{tabular}[c]{@{}l@{}} $\text{E} = 1.096677\times{C_{\text{1}}} + 2.364495\times{C_{\text{2}}} + 1.627854\times{C_{\text{3}}} + 1.173680\times{C_{\text{4}}} + 0.681475\times{C_{\text{5}}} + 0.652665\times{C_{\text{6}}}$ \end{tabular} & 3.65 & 4.08 \\ \hline
[48, ON, 1] & \begin{tabular}[c]{@{}l@{}} $\text{E} = 0.816331\times{C_{\text{1}}} + 2.014612\times{C_{\text{2}}} + 1.372157\times{C_{\text{3}}} + 1.402116\times{C_{\text{4}}} + 0.835035\times{C_{\text{5}}} + 1.250446\times{C_{\text{6}}}$ \end{tabular} & 4.33 & 4.99 \\ \hline
\end{tabular}%
}
\caption{Energy models for selected Cortex-M0 hardware configurations -- Hardware Config. Format: [Frequency (MHz), PreFetch (ON/OFF), WaitState (0/1)], MAPE: Mean Absolute Percentage Error, and RESD: Relative Error Standard Deviation}
\label{tab:models}
\end{table*}

\subsection{PMC-based Code-level Energy Modelling}

PMC-based energy consumption estimation models are typically obtained via multi-linear regression analysis, where coefficients, $\beta_x$, are determined for each counter, $C_x$, to predict the overall energy cost, i.e., $E = \sum_x (\beta_x \times C_x) + \alpha$, with $\alpha$ being the error term. The coefficients $\beta_x$ are the constants in the energy model that are program independent while the counters $C_x$ are the variables that depend on the program and its input. For a specific program with known counters, the energy model can be used to estimate the energy consumed during the program's execution.

For static-analysis-based energy consumption estimation, the overall energy consumption estimate of a piece of code is typically constructed from the estimates of the ISA basic blocks of the program \cite{Georgiou:2017}. Thus, a PMC-based energy model can enable energy consumption estimation via static analysis only if the counters used for the modeling and prediction can be statically predicted at the ISA basic block level.

\subsection{Collection of Cortex-M0 Event Counters}

Traditionally, there are two ways to collect execution statistics for an architecture. One is to collect them directly via PMCs while executing a program on the actual architecture, provided it offers PMCs. For architectures without PMCs, the second way is via an ISS, preferably cycle-accurate. The ISS simulates the execution of a program for a specific architecture, and thus, it can collect PMCs. Since the Cortex-M0 is a deeply embedded architecture with minimal resources available on-chip, it does not expose any PMCs. Thus, we modified an open-source ISS, namely \texttt{Thumbulator} \cite{Thumbulator}, to extract the necessary event counters for our energy consumption modelling. The modifications wrt.\ the reference \texttt{Thumbulator} implementation \cite{Thumbulator} included four key aspects:

\begin{itemize}
\item Adaptation to reflect the memory organisation of the STM32F0xx processor family;
\item Introduction of a model of the instruction fetch mechanism used in the STM32F0xx processors;
\item Implementation of a range of event counters and the associated reporting mechanism;
\item Calibration and improvement of the timing behaviour of the simulation to match the hardware's behaviour.
\end{itemize}

The modified simulator can be used to simulate any of the processors in the STM32F0xx family \cite{STMF32F0xx} and can collect a large number of event counters that represent various aspects of the architecture's runtime behaviour such as the effective RAM and Flash memory accesses, taken branches, per-opcode instruction execution statistics, and interactions between instruction- and data-related memory accesses. The timing behaviour validation of the modified \texttt{Thumbulator} against the actual hardware, using all the benchmarks, exposed a correctness bug in the implementation of the \texttt{ASR} instruction in the original \texttt{Thumbulator} code and identified a case of incorrect memory access counting. Both problems have been fixed in the version used to build the final energy model. The execution time model derived from event counts reported by \texttt{Thumbulator} is fully cycle-accurate wrt. hardware execution when the instruction PreFetch buffer is disabled or the WaitState count is 0. When the PreFetch buffer is enabled and the WaitState count is 1, the MAPE of the \texttt{Thumbulator}-based timing prediction is 1.55\%.

Using the available architecture documentation and a series of modeling cycles, we constrained the number of event counters used for the modeling to the set of the counters that have the most significant impact on the energy consumption and are suitable for static analysis. These counters also yield the highest observed estimation accuracy compared to physical measurements when compared with the retrieved estimations of other event counter combinations. The selected counters are shown in \Cref{tab:Cortex_M0_counters}.

\subsection{Model Training and Validation}
\label{subsec:Cortex_m0_model_training}

Both the BEEBS and CNN-based benchmarks have been compiled into two kinds of binaries. First, the benchmarks have been compiled for the STM32F0-DISCOVERY board in order to conduct energy consumption measurements. The hardware measured energy consumption of the programs provide the data for the dependent variables of our regression analysis. Second, the benchmarks have been compiled for the modified \texttt{Thumbulator} ISS in order to derive events counter values. The counter values provide the data for the independent variables of our regression analysis. The two sets of binaries are required because the simulator does not fully handle access to off-core peripherals, e.g., PLL clock generators; these should be skipped in \texttt{Thumbulator} binaries. However, the same location and alignment of benchmark code for both types of binaries was maintained. 

\begin{figure}[!htb]
\centering
\resizebox{1\textwidth}{!}{%
  \includegraphics[trim=0.5cm 12cm 6.2cm 1.6cm,clip=true,width=0.5\linewidth]{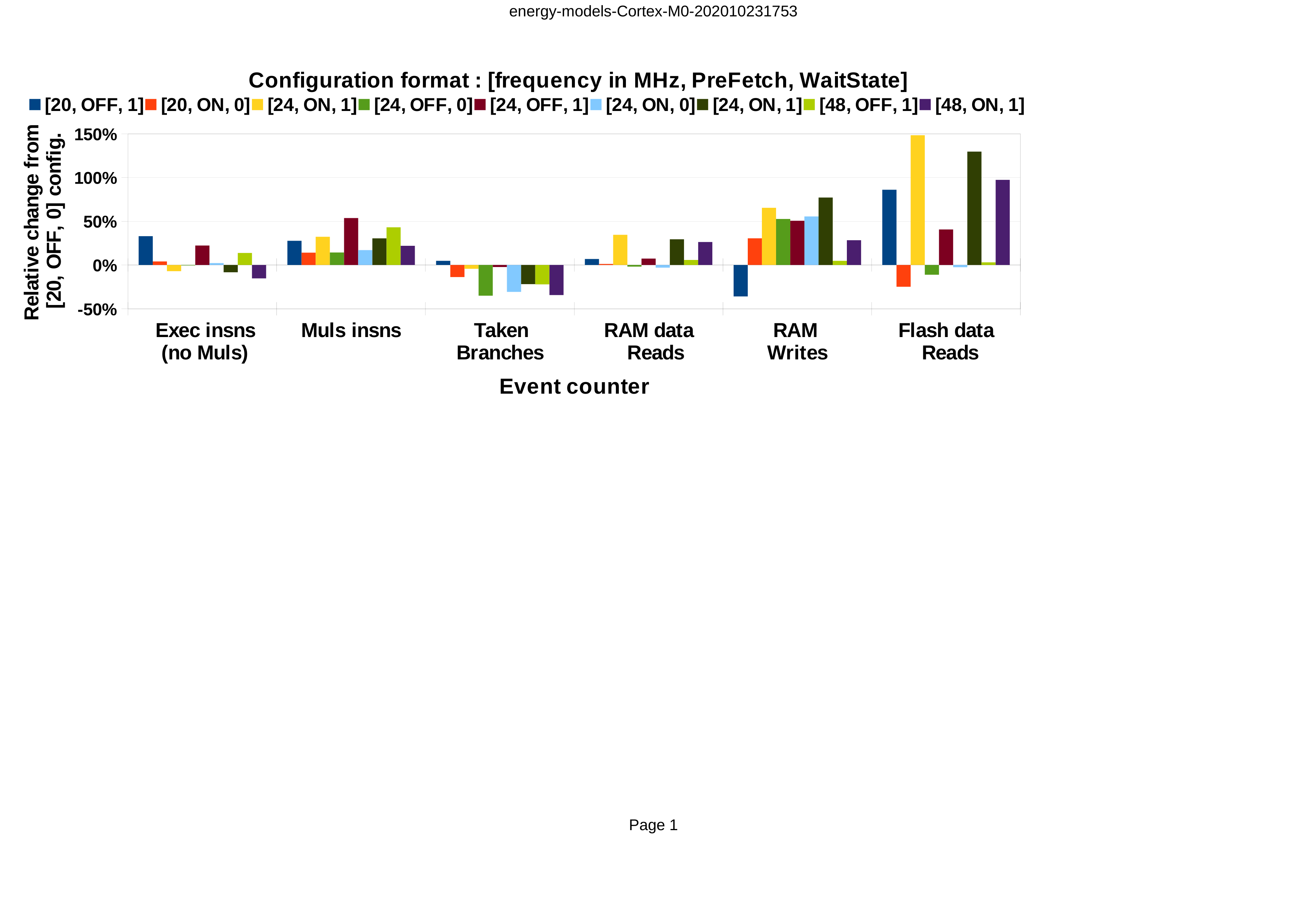}
}
\caption{Relative costs associated with events used in the energy model at distinct processor configurations.}
\label{fig:relativeEventWeigthAnalysis}
\end{figure}

When using regression modelling, it is critical to include as broad and representative a training sample as possible in the training phase. This ensures that the model is as generic as possible and can capture a large part of the space being modeled. Thus, instead of splitting our data into predefined training and testing sets, we included all data into the training, and we used k-fold cross-validation to ensure the retrieved models avoid overfitting and selection bias. If the cross-validation demonstrates a good estimation accuracy across all folds, then the final model using all the available data will exhibit a balanced variance and bias. Thus, the model will have a good chance of accurately capturing a big part of the space being modeled. In our case, we used 10-fold cross-validation and we used the $R^2$ to evaluate the performance of each of the ten models for each of the modelling configuration, shown in \Cref{tab:models} under column \emph{Hardware Configuration}. An $R^2$ value close to 1 demonstrates an excellent prediction. The 10-fold cross-validation yielded an $R^2$ mean value of close to 0.99 for all configurations, with a standard deviation of around 0.2\%. This is a robust result as the $R^2$ score approaches the value of one across all the different folds, demonstrating that the counters selected for the model are accurately capturing the energy consumption of a variety of programs. Thus, for the final model, all the data points were used for training.
\label{subsec:security}

Energy models for the different hardware configurations and their accuracy are listed in \Cref{tab:models}. For all models the MAPE is less than 5\%, with a standard deviation of less than 5\%, compared to hardware energy measurements. An early version of the model, configured for 20 MHz frequency, PreFetch on, and WaitState 1 was evaluated in the context of static energy consumption estimation in~\cite{D4.4_TeamPlay} demonstrating the suitability of our models for static analysis. 

\subsection{Potential Use-Case in Cyber Attacks}

A comparative analysis of energy model coefficients extracted at distinct processor configurations (see~\Cref{fig:relativeEventWeigthAnalysis}) shows significant variation in relative weights of the different events across the hardware configurations. It follows that by comparing the energy consumption of the processor at distinct frequency, PreFetch, and WaitState settings and by applying statistical model fitting techniques, an observer can potentially predict the proportion of each event in the program.  Subsequently, the observer will also be able to predict the type of processing being performed (e.g., data- vs. control-centric). By increasing the time resolution of analysis (narrowing the observed time window), the observer could also identify distinct program execution phases. From a security standpoint, such information leakage forms a potential side-channel for attacks. These attacks can be directed against the secret information contained in the program code (e.g., encryption keys or ciphering algorithms) or against the data being processed, e.g., in autonomous medical diagnostics devices.

\section{Conclusion and future work}

This paper offers an open-source, ready-to-use energy model for the Arm Cortex-M0 processor. The model can be used for profiling-based analysis to estimate the actual energy consumption, and in static analysis to estimate the energy consumed by the worst-case execution path of an edge-computing application. 
Furthermore, the model accounts for the frequency and the flash instruction-buffer configurations of the processor that can significantly affect the execution time and energy consumption of an application, namely, PreFetch, and WaitState instruction-buffer configurations. Our customized open-source ISS is also readily available to profile the execution time and energy consumption of edge computing applications for any of the STM32F0xx family of processors. This allows developers to choose the hardware configuration that can meet the resource requirements. Preliminary analysis indicates that our models can be exploited for side-channel attacks to reveal information about the type of application and the processing an application performs. Future work will test these observations. 

\section*{Acknowledgments}
This research is supported by the European Union's Horizon 2020 Research and Innovation Programme under grant agreement No. 779882, TeamPlay (Time, Energy and security Analysis for Multi/Many-core heterogeneous PLAtforms).

\bibliographystyle{alphaurl}
\bibliography{typeinst}

\end{document}